\documentclass[aps,prc,preprint,groupedaddress]{revtex4-1}

\usepackage{graphicx}
\usepackage{amsmath,bm,ascmac,amssymb,calc}
\usepackage{CJK}

\begin{document}
\begin{CJK*}{UTF8}{}
\CJKfamily{min}


\title{Assessment of effect of local approximation on single folding potential
  at low and intermediate incident energies}


\author{K. Ishida (石田 佳香)$^1$}
\author{H. Nakada (中田 仁)$^{2,3}$}
\email[E-mail:\,\,]{nakada@faculty.chiba-u.jp}

\affiliation{$^1$ Department of Physics,
 Graduate School of Science and Engineering,
 Chiba University,\\
Yayoi-cho 1-33, Inage, Chiba 263-8522, Japan}
\affiliation{$^2$ Department of Physics, Graduate School of Science,
 Chiba University,\\
Yayoi-cho 1-33, Inage, Chiba 263-8522, Japan}
\affiliation{$^3$ Research Center for Nuclear Physics, Osaka University,\\
  Mihogaoka 10-1, Ibaraki, Osaka 567-0047, Japan}


\date{\today}

\begin{abstract}
  To the single folding potentials (SFPs)
  for the nucleon-nucleus ($N$-$A$) elastic scatterings,
  local approximations (LAs) have customarily been applied.
  The LA discussed by Brieva and Rook has been well-known,
  which only needs the density profile as the structure information
  of the target nucleus.
  By applying the M3Y-P6 interaction
  both to the target wave functions and the real part of SFP,
  supplemented with the Koning-Delaroche phenomenological imaginary potential,
  the precision of the Brieva-Rook LA on the SFP is investigated
  for the proton-nucleus elastic scatterings
  at $\epsilon_p=16\,-\,80\,\mathrm{MeV}$ incident energies.
  The analyzing powers as well as the differential cross sections
  are in reasonable agreement with the available data.
  The precision of the LA for the central and LS channels
  is distinctly examined.
  Although the LA works well at small angles
  ($\theta_\mathrm{c.m.}\lesssim 30^\circ$),
  it gives rise to sizable deviation from the results of the non-local SFP
  (\textit{i.e.}, without the LA) at larger angles.
  The results of the non-local SFP are always in better agreement with the data.
  The LA for the LS channel influences the differential cross-sections,
  and the LA for the central channel does the spin observables.
  It is found that the precision of the LA well correlates
  to the momentum transfer $q$,
  and the discrepancy becomes sizable at $q\gtrsim 1.5\,\mathrm{fm}^{-1}$.
  The LA is also examined for a halo nucleus,
  by taking $^{86}$Ni as an example.
  The precision is slightly worse than in stable nuclei.
  Difference from the prediction of the empirical potential
  in the observables of the $p$-$^{86}$Ni scattering is discussed.
\end{abstract}


\maketitle
\end{CJK*}



\section{Introduction\label{sec:intro}}

The optical potential supplies a framework
describing the elastic scatterings of composite objects,
being indispensable
to analyzing low-energy nuclear reactions~\cite{ref:Gle83,ref:Sat83}.
The optical potential is given as
\begin{equation}
 \mathcal{U} = \mathcal{V} + i\,\mathcal{W}\,,
\label{eq:OMP}\end{equation}
where $\mathcal{V}$ and $\mathcal{W}$ are one-body hermitian operators.
The nuclear optical potential was conventionally obtained
by assuming $\mathcal{V}$ and $\mathcal{W}$ to be local operators
and fitting their parameters to the experimental data.
The optical potential can be derived from the nucleonic interaction
as the folding potential~\cite{ref:SL79}.
However, although the folding potential should have non-locality in general,
\textit{e.g.}, the non-locality arising from the exchange term
of the nucleonic interaction,
practical calculations using the folding potential have been implemented
under local approximations (LAs)
because of the computation time and the available computer codes.
In particular, the LA formulated by Brieva and Rook
in Refs.~\cite{ref:BR77a,ref:BR77b,ref:BR78}
has widely been applied.

The nucleon-nucleus ($N$-$A$) scattering is fundamental
to low- and intermediate-energy nuclear reactions,
since the compositeness of target $A$ is relevant
but that of the projectile $N$ is not.
The systematics of the parameters
with respect to the incident energy $\epsilon$ and the mass number $A$
have been analyzed for the empirical optical potential~\cite{ref:CH89,ref:KD03}.
On the other hand, 
the single-folding potential (SFP) has been calculated and applied
via the folding for target $A$ with nucleonic effective interactions.
While the SFP generally contains the one-body density matrix (DM)
of the target nucleus,
the nucleon density distribution $\rho_\tau(r)$ ($\tau=p,n$),
the diagonal elements of the DM in the coordinate representation,
is the only nuclear structure information needed in the SFP
under the LA of Refs.~\cite{ref:BR77a,ref:BR77b,ref:BR78}.
Therefore, the scattering observables are well connected to the density profile,
as far as this LA is precise.
In contrast,
we need information on the target wave function beyond $\rho_\tau(r)$
for calculations of the SFP without the LA.
Though vital to correctly extract nuclear structure information
from the scattering data,
the precision of the LA has not sufficiently been assessed.
The differential cross-sections with and without the LA were compared
in Ref.~\cite{ref:LKP20},
limiting the non-locality in the central channel of the nucleonic interaction
in $\epsilon<50\,\mathrm{MeV}$.
The wave functions of the projectile and target
were calculated with different interactions,
obscuring whether the non-local SFP provides a rational baseline.
In Ref.~\cite{ref:ABL90},
the precision of LAs was investigated
for the $p$-$^{16}$O and $p$-$^{40}$Ca scatterings
at relatively high energies $\epsilon_p=200\,-\,400\,\mathrm{MeV}$,
employing the $t$-matrix.
The wave function of the constituent nucleons in the target
was obtained from the Woods-Saxon (WS) potential.
Few arguments were provided for the $\ell s$ potential.
It was claimed that the LA is valid at $q\lesssim 2.5\,\mathrm{fm}^{-1}$,
where $q$ denotes the momentum transfer.
Whereas this range of $q$ covers the whole range of the scattering angles
at low energies,
many-body correlations beyond this calculation
might not be negligible at low energies.
We mention that the so-called $JvH$ factorization scheme,
which is beyond the LA but keeps a relatively simple structure in the SFP,
has recently been proposed and applied
up to $400\,\mathrm{MeV}$ incident energy~\cite{ref:AB24},

The Michigan-three-range-Yukawa (M3Y) effective interaction
was developed based on the $G$-matrix~\cite{ref:M3Y,ref:M3Y-P}.
The M3Y interaction was extended by introducing density-dependent terms,
so that it could apply to the nuclear structure~\cite{ref:Nak03}.
In particular, the M3Y-P6 interaction~\cite{ref:Nak13}
has been found to describe the nuclear shell structure
up to the $Z$- and $N$-dependence~\cite{ref:NS14,ref:Nak20},
indicating its reliability for the single-particle (s.p.) potential
at negative energies.
We have extensively applied the M3Y-P6 interaction
to the $N$-$A$ elastic scatterings in Ref.~\cite{ref:NI24},
by computing both the projectile and target wave functions
with this interaction,
except for the imaginary part of the optical potential.
With good s.p. potentials continuous in energy,
we now have a good opportunity to investigate the precision of the LA
at low and intermediate energies,
owing to the development of computer codes~\cite{ref:SIDES}
and to the formulation and the reasonable results
of the non-local SFP under the consistent interaction.

\section{Single folding potential
  with and without local approximation\label{sec:SFP}}

\subsection{Effective Hamiltonian\label{subsec:eff-H}}

For obtaining the folding potential microscopically,
the $G$-matrix in the momentum representation supplies a suitable base.
It is often converted to the effective interaction
represented in terms of the relative coordinate of two nucleons~\cite{ref:RKG84,
  ref:YNM86,ref:ADG00}.
However, complicated many-body correlations play roles at low energies,
as typically found in nuclear structure problems.
A framework to handle the many-body correlations
is given by the Kohn-Sham (KS) theory~\cite{ref:KS65},
as primarily discussed for ground-state properties.
Most generally,
the KS theory relies on the variation of the energy functional
with respect to the one-body DM~\cite{ref:Nak23}.
The energy functional in the KS theory can be associated
with the nucleonic effective interaction.
In Ref.~~\cite{ref:NI24},
the SFP has been formulated via the variation of the energy functional,
consistently with the KS approach,
and an appropriate energy functional (or effective interaction)
has been shown to be applicable from the nuclear structure
to the scattering at intermediate incident energies,
evidencing good continuity in energy.
It is commented that the variational derivation leads to
the density rearrangement term~\cite{ref:LKP20,ref:NS06},
which was ignored
in the conventional folding potentials~\cite{ref:RKG84,ref:YNM86,ref:ADG00}.

In the following, we consider an effective Hamiltonian,
\begin{equation}\begin{split} H =& K + V_N + V_C - H_\mathrm{c.m.}\,;\\
& K = \sum_\alpha \frac{\mathbf{p}_\alpha^2}{2M}\,,\quad
V_N = \sum_{\alpha<\beta} v_{\alpha\beta}\,,\quad
V_C = \alpha_\mathrm{em} \sum_{\alpha<\beta(\in p)} \frac{1}{r_{\alpha\beta}}\,,\\
& H_\mathrm{c.m.} = \frac{\mathbf{P}^2}{2A'M}
= \frac{1}{A'}\bigg[\sum_\alpha \frac{\mathbf{p}_\alpha^2}{2M}
  + \sum_{\alpha<\beta} \frac{\mathbf{p}_\alpha\cdot\mathbf{p}_\beta}{M}\bigg]\,,
\end{split}\label{eq:Hamil}\end{equation}
where the subscripts $\alpha$ and $\beta$ are nucleons' indices,
$\mathbf{r}_{\alpha\beta}= \mathbf{r}_\alpha - \mathbf{r}_\beta$,
$r=|\mathbf{r}|$, $\mathbf{P}=\sum_\alpha \mathbf{p}_\alpha$,
$A'$ is the mass number of the target,
and $\alpha_\mathrm{em}$ (in $V_C$) denotes the fine structure constant.
The two-nucleon interaction $v_{\alpha\beta}$ in $V_N$
is comprised of the central, LS and tensor channels,
\begin{equation} v_{\alpha\beta} = v_{\alpha\beta}^{(\mathrm{C})}
 + v_{\alpha\beta}^{(\mathrm{LS})} + v_{\alpha\beta}^{(\mathrm{TN})}
 + v_{\alpha\beta}^{(\mathrm{C}\rho)}\,.
 \label{eq:effint}\end{equation}
The density-dependent contact term of the central channel
$v_{\alpha\beta}^{(\mathrm{C}\rho)}$ is responsible for the saturation.
All terms except for $v_{\alpha\beta}^{(\mathrm{C}\rho)}$
have finite-range Yukawa functions of $r_{\alpha\beta}$
in the M3Y-type interaction~\cite{ref:Nak03}.

\subsection{Non-local single folding potential\label{subsec:SFP}}

Suppose that the total energy of system $E$ is represented
in terms of the DM
as in the self-consistent mean-field (SCMF) or the KS approaches.
The s.p. Hamiltonian $h$ is then derived
via the variation of $E$ with respect to the DM,
defining the s.p. potential $U$ by
\begin{equation}
  h = \frac{\mathbf{p}^2}{2M} + U\,.
  \label{eq:sp-hamil}\end{equation}
This $U$ can be identified as $\mathcal{V}$ of \eqref{eq:OMP}
at positive energies~\cite{ref:NI24}.
Unless the effective interaction is constrained to have zero range,
the exchange term leads to non-locality in the SFP.
The formulae needed to calculate the non-local SFP
have been provided in Appendices of Ref.~\cite{ref:NI24}.
We denote $\mathcal{V}$ by $\mathcal{V}^\mathrm{SFP}$
when it is obtained by the variation of $E$ with Eq.~\eqref{eq:sp-hamil}.

\subsection{Local approximation on single folding potential
  \label{subsec:LocApp}}

We investigate the LA on the SFP
formulated in Refs.~\cite{ref:BR77a,ref:BR77b,ref:BR78},
which have been popular in practical applications of the SFP.
To distinguish from the non-local SFP,
we denote the approximated potential by $\tilde{\mathcal{V}}^\mathrm{SFP}$,
which is composed of the central and spin-orbit ($\ell s$) terms,
$\tilde{\mathcal{V}}^{(\mathrm{ct})}$ and $\tilde{\mathcal{V}}^{(\ell s)}$.

In the LA of Ref.~\cite{ref:BR77b},
the Slater approximation was applied to the DM
for the SFP from the central channel.
In Ref.~\cite{ref:ABL90},
a LA obtained from the Campi-Bouyssy (CB) approximation
on the DM~\cite{ref:CB78}
was compared with the LA with the Slater approximation.
While the CB approximation needs the kinetic density of the target,
which is nuclear structure information beyond $\rho_\tau(r)$,
its influence is insignificant.
Taking the 1st two terms of the DM expansion,
the Negele-Vautherin (NV)~\cite{ref:NV72} approximation
supplies another LA,
which also depends on the kinetic density.
The NV approximation
does not give significant difference from the Slater approximation
for the scattering observables, either.
LAs have not been explored sufficiently
for the spin-dependent parts.
It has been shown that the NV expansion is less precise
for the spin-dependent channel~\cite{ref:DCK10}.
Not many discussions have been provided for the LS channel,
while the LA of Ref.~\cite{ref:BR78} is well-known.
In the following, we shall focus on the LA
of Refs.~\cite{ref:BR77a,ref:BR77b,ref:BR78},
which is summarized in Appendix.
Note that the full $\mathcal{V}^\mathrm{SFP}$ depends on $\ell$ and $j$
(the orbital and summed angular momenta
of the scattered nucleon)~\cite{ref:NI24},
not simply separable into central and $\ell s$ terms.

Under the LA,
the central channel of the nucleonic interaction,
$v_{\alpha\beta}^{(\mathrm{C})}+v_{\alpha\beta}^{(\mathrm{C}\rho)}$,
determines $\tilde{\mathcal{V}}^{(\mathrm{ct})}$.
The LS channel $v_{\alpha\beta}^{(\mathrm{LS})}$
derives $\tilde{\mathcal{V}}^{(\ell s)}$.
Although the zero-range form of the LS interaction
is adopted in some effective interactions~\cite{ref:VB72,ref:Gogny},
on which LA has no effects,
it is not sufficient to describe observed nuclear properties,
\textit{e.g.}, the kink in the isotopic difference
of the nuclear charge radii~\cite{ref:SLKR95,ref:NI15,ref:Nak15}.
The full formulation of $\mathcal{V}^\mathrm{SFP}$ given in Ref.~\cite{ref:NI24}
enables the assessment of the LA for $v_{\alpha\beta}^{(\mathrm{LS})}$.
To distinguish the influence of the LA on $v_{\alpha\beta}^{(\mathrm{LS})}$
from that on $v_{\alpha\beta}^{(\mathrm{C})}$,
we also consider a potential
$\tilde{\mathcal{V}}^{\mathrm{SFP}(\tilde{\mathrm{C}}+\mathrm{LS})}$,
in which the LA is applied to $v_{\alpha\beta}^{(\mathrm{C})}$
but not to $v_{\alpha\beta}^{(\mathrm{LS})}$.
The LA for $v_{\alpha\beta}^{(\mathrm{C})}$
can then be assessed from the difference
between the results of $\mathcal{V}^\mathrm{SFP}$
and those of $\tilde{\mathcal{V}}^{\mathrm{SFP}(\tilde{\mathrm{C}}+\mathrm{LS})}$,
and the LA for $v_{\alpha\beta}^{(\mathrm{LS})}$ from the difference
between $\tilde{\mathcal{V}}^{\mathrm{SFP}(\tilde{\mathrm{C}}+\mathrm{LS})}$
and $\tilde{\mathcal{V}}^\mathrm{SFP}$.
We confirm that effects of the tensor force $v_{\alpha\beta}^{(\mathrm{TN})}$
on the scattering observables are not visible,
except for the spin rotation at limited angles in the halo nucleus.

\section{Numerical calculations\label{sec:cal}}

\subsection{Numerical setups\label{subsec:setup}}

This paper focuses on the proton scatterings,
on which abundant data are available.
While the neutron scatterings have been calculated,
they do not influence the arguments below.
We treat $^{16}$O, $^{40}$Ca, $^{90}$Zr and $^{208}$Pb as target nuclei,
ranging from light- to heavy-mass regions.
Since they are doubly magic nuclei,
their ground-state wave functions are well described
in the spherical Hartree-Fock (HF) calculations.
It should be recalled that the densities or radii of the targets,
to which the scattering observables have relevance,
have been examined in Refs.~\cite{ref:Nak13,ref:Nak19}.
In addition, results of the proton scattering
off a highly neutron-rich nucleus $^{86}$Ni will be shown,
which may also be a doubly magic nucleus~\cite{ref:Nak10}
and is predicted to have a neutron halo (see Sec.~\ref{subsec:Ni86}).
We cover the incident energies as broad as
$\epsilon_p=16$\,--\,$80\,\mathrm{MeV}$,
where M3Y-P6 reproduces the measured differential cross-sections
reasonably well~\cite{ref:NI24}.
Relativistic effects may partly be incorporated
into the effective interaction in this energy range.
In all cases,
we apply the same M3Y-P6 interaction
both for the HF calculations of the target
and the calculations of $\mathcal{V}^\mathrm{SFP}$.

The imaginary potential $\mathcal{W}$ represents
the influence of the virtual excitations of the target.
Whereas it can also be non-local,
the non-locality in $\mathcal{W}$ has not sufficiently been understood.
There should be two sources to induce the imaginary potential.
The imaginary part arises in the effective nucleonic interaction
in the Brueckner theory,
originating primarily from the high-momentum component
of the bare nucleonic interaction.
The imaginary part of the interaction yields
the imaginary part of the optical potential through the folding~\cite{ref:RKG84,
  ref:YNM86,ref:ADG00,ref:JLM76,ref:FSY08,ref:HKMW13,ref:VFG16,ref:WLH21}.
On the other hand,
the imaginary potential may arise from the collective low-momentum excitations,
as can be treated via the particle-vibration coupling picture~\cite{ref:MO12,
  ref:Bla15}.
It is not yet possible to derive imaginary potentials
by taking into account both components simultaneously.
We employ the empirical imaginary potential of Ref.~\cite{ref:KD03}
as in Ref.~\cite{ref:NI24},
which is local and energy-dependent.

The scattering wave is obtained
by numerically solving the integro-differential Schr\"{o}dinger equation
via the SIDES code~\cite{ref:SIDES}.
We need the input parameters $R$, $\mathit{\Delta}r$,
and $\ell_\mathrm{max}$ parameters:
the maximum radius, the radial mesh,
and the maximum partial wave.
As the effective interaction has no energy dependence,
the non-local $\mathcal{V}^\mathrm{SFP}$ is energy-independent,
whereas the LA brings the dependence on the incident energy
into $\tilde{\mathcal{V}}^\mathrm{SFP}$.
Since the non-local $\mathcal{V}^\mathrm{SFP}$ applies
to any energies once calculated,
it is convenient to employ a single set
of the $R$, $\mathit{\Delta}r$, and $\ell_\mathrm{max}$ parameters,
independent of the incident energies.
In this paper,
$R=15\,\mathrm{fm}$, $\mathit{\Delta}r=0.02\,\mathrm{fm}$
and $\ell_\mathrm{max}=30$ are adopted,
after confirming the convergence for all the cases handled in this paper.
The scattering wave beyond $R$ is continuated to the asymptotic form.

\subsection{Differential cross-sections\label{subsec:DCS}}

\begin{figure}
  \hspace*{-4cm}\includegraphics[scale=0.5]{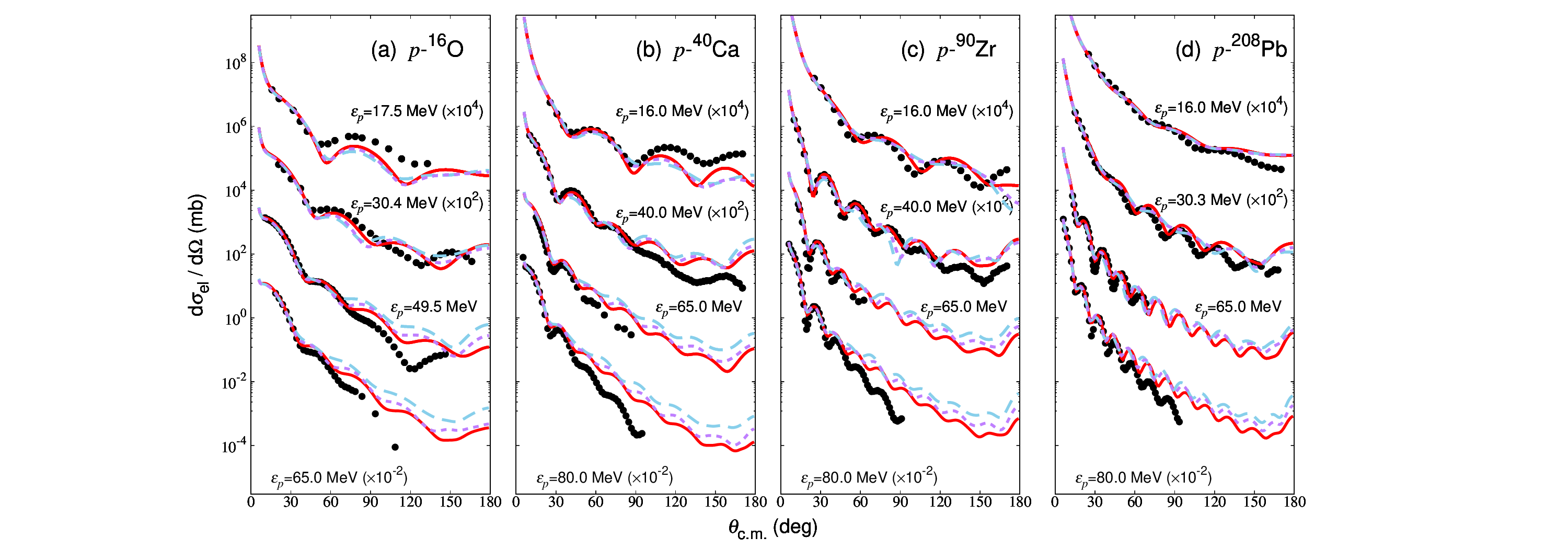}
\caption{Differential cross-sections of $p$-$A$ elastic scatterings:
  (a) $p$-$^{16}$O at $\epsilon_p=17.5$, $30.4$, $49.5$ and $65.0\,\mathrm{MeV}$,
  (b) $p$-$^{40}$Ca at $\epsilon_p=16.0$, $40.0$, $65.0$ and $80.0\,\mathrm{MeV}$,
  (c) $p$-$^{90}$Zr at $\epsilon_p=16.0$, $40.0$, $65.0$ and $80.0\,\mathrm{MeV}$,
  (d) $p$-$^{208}$Pb at $\epsilon_p=16.0$, $30.3$, $65.0$ and $80.0\,\mathrm{MeV}$.
  Results of $\mathcal{V}^\mathrm{SFP}$,
  $\tilde{\mathcal{V}}^{\mathrm{SFP}(\tilde{\mathrm{C}}+\mathrm{LS})}$
  and $\tilde{\mathcal{V}}^\mathrm{SFP}$
  are depicted by red solid, purple dotted and skyblue dashed lines,
  respectively.
  For comparison, experimental data taken from the database~\cite{ref:EXFOR}
  are shown by black circles.
  These data were originally reported
  in Refs.~\protect\cite{ref:CH89,ref:exp_DCS_RCNP,ref:exp_DCS_p-O16_e17,
        ref:exp_DCS_p-O16_e30,ref:exp_DCS_p-O16_e50,ref:exp_DCS_p-A_e40,
        ref:exp_DCS_p-A_e80,ref:exp_DCS_p-Pb208}.
  Depending on $\epsilon_p$,
  the values are scaled by the coefficient given in the parenthesis.
\label{fig:p-A_dsig}}
\end{figure}

The results of the differential cross sections $d\sigma_\mathrm{el}/d\Omega$
are shown in Fig.~\ref{fig:p-A_dsig}.
As has been shown in Ref.~\cite{ref:NI24},
the non-local SFP (\textit{i.e.}, $\mathcal{V}^\mathrm{SFP}$),
to which the M3Y-P6 interaction is applied
consistently with the target wave functions,
reproduces the experimental data reasonably well.
By comparing the results of $\mathcal{V}^\mathrm{SFP}$
and $\tilde{\mathcal{V}}^\mathrm{SFP}$,
we find that the LA works excellently at $\theta_\mathrm{c.m.}\lesssim 30^\circ$,
irrespective of the target and the incident energy.
However, its precision is not very high at larger scattering angles,
except for the $p$-$^{208}$Pb scattering at $\epsilon_p=16\,\mathrm{MeV}$.
The results of $\tilde{\mathcal{V}}^\mathrm{SFP}$
(\textit{i.e.}, the results with the LA)
deviate from those of $\mathcal{V}^\mathrm{SFP}$
(\textit{i.e.}, the results without the LA)
at $\theta_\mathrm{c.m.}\gtrsim 50^\circ$.
Although the results of $\mathcal{V}^\mathrm{SFP}$
and $\tilde{\mathcal{V}}^\mathrm{SFP}$
tend to be close at lower energies,
the discrepancy is already visible at $\epsilon_p=40\,\mathrm{MeV}$,
not strongly depending on $A$.
Even positions of the peaks and dips shift.
It should be noted
that $\mathcal{V}^\mathrm{SFP}$ describes the measured cross section
better than $\tilde{\mathcal{V}}^\mathrm{SFP}$
in all the cases under investigation.
From the results
of $\tilde{\mathcal{V}}^{\mathrm{SFP}(\tilde{\mathrm{C}}+\mathrm{LS})}$,
we find that the LA for $v_{\alpha\beta}^{(\mathrm{LS})}$
influences the differential cross-sections
even more significantly than the LA for $v_{\alpha\beta}^{(\mathrm{C})}$.

\begin{figure}
  \hspace*{2cm}\includegraphics[scale=0.5]{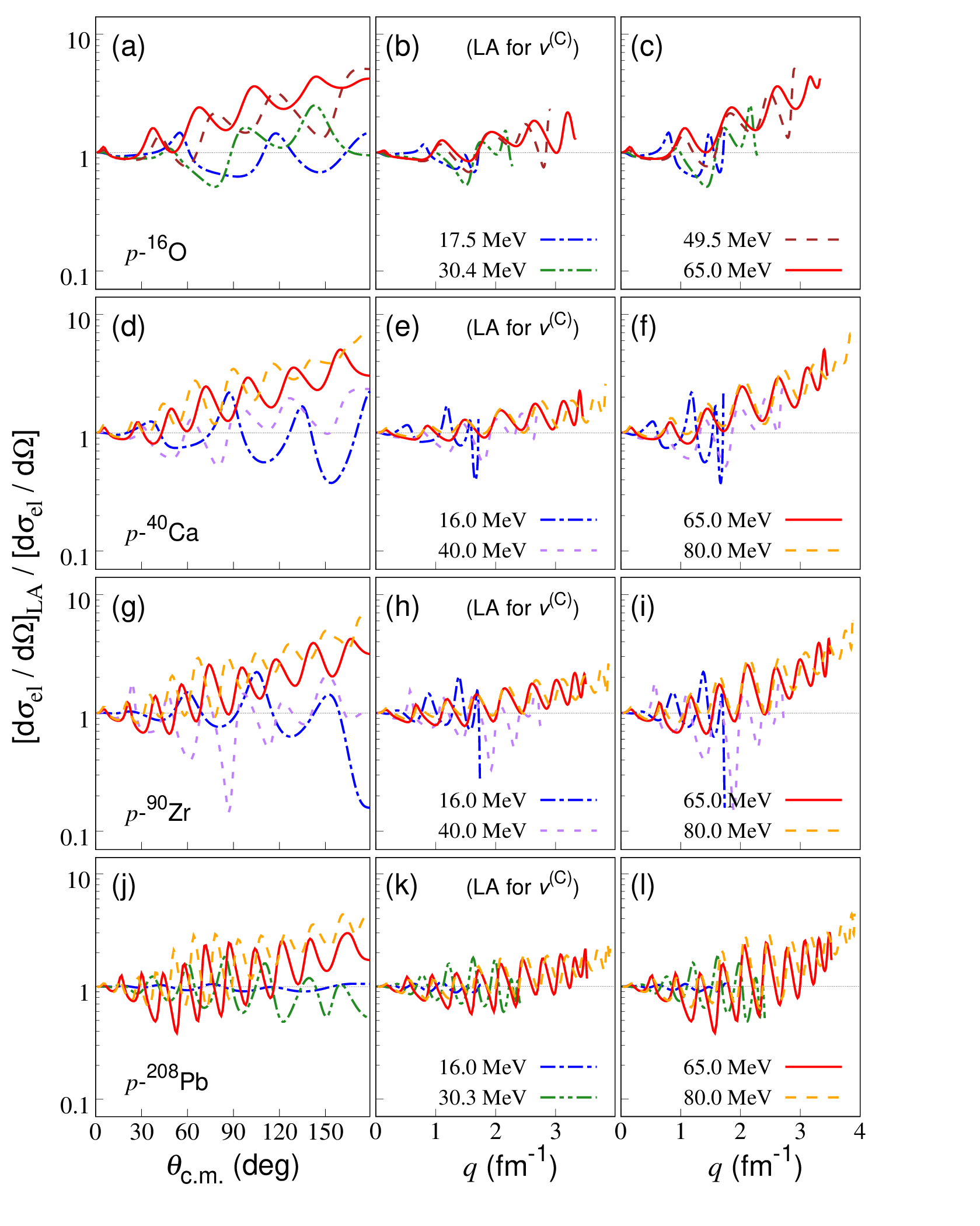}
  \caption{Ratios of differential cross-sections with LA to those without LA
    for the $p$-$A$ elastic scatterings,
    as functions of $\theta_\mathrm{c.m.}$ [left panels (a,d,g,j)]
    and $q$ [middle and right panels (the others)]:
    (a,b,c) $p$-$^{16}$O, (d,e,f) $p$-$^{40}$Ca,
    (g,h,i) $p$-$^{90}$Zr, (j,k,l) $p$-$^{208}$Pb.
    Incident energies for the individual target
    are distinguished by the line types as indicated in the panels.
    The left [(a,d,g,j)] and right [(c,f,i,l)] panels depict the ratios
    of the cross-sections with $\tilde{\mathcal{V}}^{\mathrm{SFP}}$
    to those with $\mathcal{V}^\mathrm{SFP}$,
    while the middle [(b,e,h,k)] panels the ratios of the cross-sections
    with $\tilde{\mathcal{V}}^{\mathrm{SFP}(\tilde{\mathrm{C}}+\mathrm{LS})}$
    to those with $\mathcal{V}^\mathrm{SFP}$.
\label{fig:p-A_dsig-err}}
\end{figure}

In order to assess the LA further,
the ratios of the $\tilde{\mathcal{V}}^\mathrm{SFP}$ results
to the $\mathcal{V}^\mathrm{SFP}$ results
are depicted in Fig.~\ref{fig:p-A_dsig-err}.
They are plotted as functions of the scattering angles $\theta_\mathrm{c.m.}$
and of the momentum transfer $q$.
The slight shifts of peaks and dips observed in Fig.~\ref{fig:p-A_dsig}
give rise to the oscillating behavior.
Apart from it, we find that the precision of the LA better correlates to $q$
than $\theta_\mathrm{c.m.}$.
The LA tends to overestimate the cross sections
at $q\gtrsim 1.5\,\mathrm{fm}^{-1}$,
and the overestimation becomes more serious as $q$ grows,
irrespective of the nuclides.
The ratios of the results
of $\tilde{\mathcal{V}}^{\mathrm{SFP}(\tilde{\mathrm{C}}+\mathrm{LS})}$
to those of $\mathcal{V}^\mathrm{SFP}$
are also presented as functions of $q$.
It is confirmed that the LA for $v_{\alpha\beta}^{(\mathrm{LS})}$
influences significantly,
although the LA only for $v_{\alpha\beta}^{(\mathrm{C})}$
is moderately good even at $q\approx 2\,\mathrm{fm}^{-1}$,
as observed in Fig.~\ref{fig:p-A_dsig-err}\,(b,e,h,k),
reminding us of the consequence in Ref.~\cite{ref:ABL90}.

\subsection{Spin observables\label{subsec:Ay&Q}}

Though postponed in Ref.~\cite{ref:NI24},
we here discuss the application of the SFP with M3Y-P6
to the spin observables.

\begin{figure}
  \includegraphics[scale=0.7]{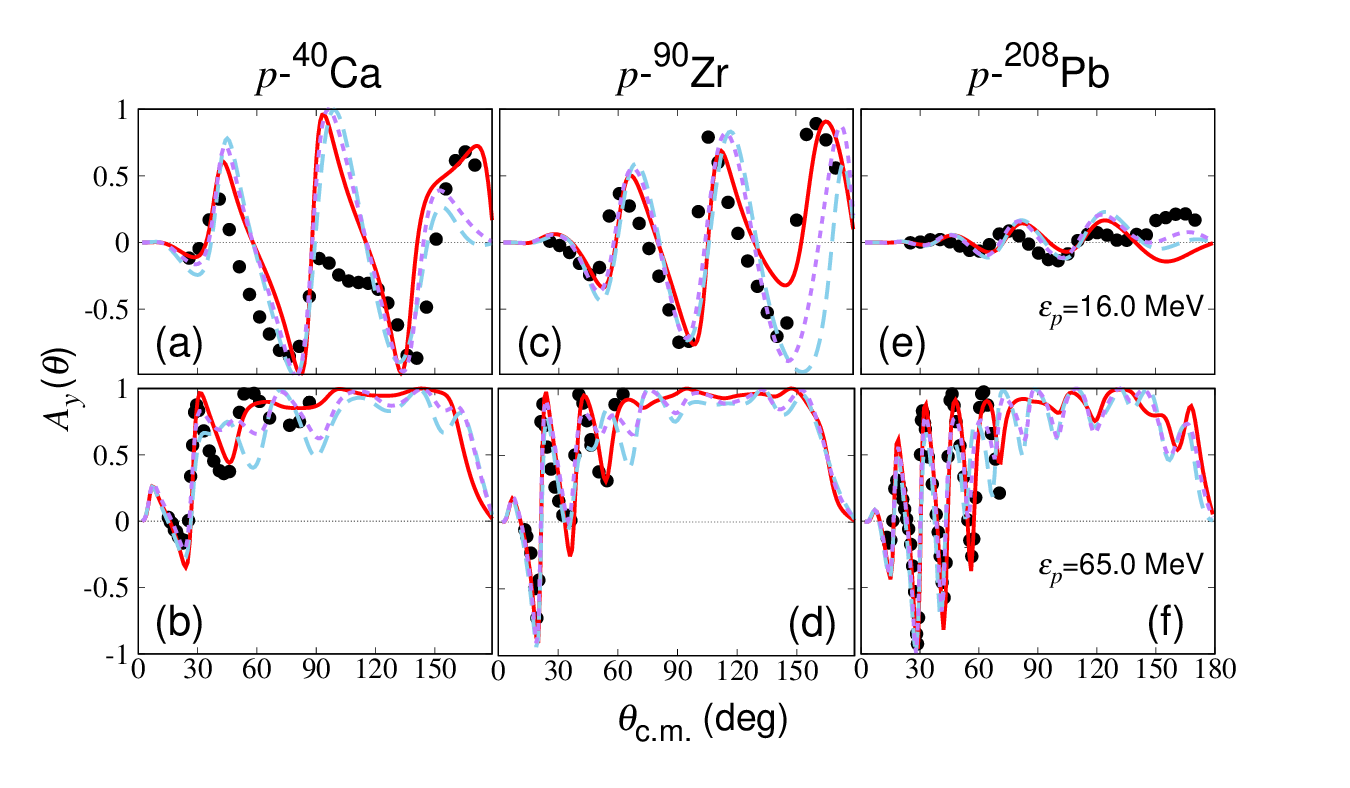}
\caption{Analyzing powers of $p$-$A$ elastic scatterings:
  (a) $p$-$^{40}$Ca at $\epsilon_p=16.0\,\mathrm{MeV}$,
  (b) $p$-$^{40}$Ca at $\epsilon_p=65.0\,\mathrm{MeV}$,
  (c) $p$-$^{90}$Zr at $\epsilon_p=16.0\,\mathrm{MeV}$,
  (d) $p$-$^{90}$Zr at $\epsilon_p=65.0\,\mathrm{MeV}$,
  (e) $p$-$^{208}$Pb at $\epsilon_p=16.0\,\mathrm{MeV}$,
  (f) $p$-$^{208}$Pb at $\epsilon_p=65.0\,\mathrm{MeV}$.
  Results of $\mathcal{V}^\mathrm{SFP}$,
  $\tilde{\mathcal{V}}^{\mathrm{SFP}(\tilde{\mathrm{C}}+\mathrm{LS})}$
  and $\tilde{\mathcal{V}}^\mathrm{SFP}$
  are depicted by red solid, purple dotted and skyblue dashed lines,
  respectively.
  Experimental data (black circles) taken from the database~\cite{ref:EXFOR}
  are also presented for comparison.
  These data were originally reported
  in Refs.~\protect\cite{ref:CH89,ref:exp_DCS_RCNP}.
\label{fig:p-A_Ay}}
\end{figure}

A number of experimental data are available for analyzing power.
The analyzing power is primarily subject
to the LS channel of the nucleonic interaction.
In Fig.~\ref{fig:p-A_Ay},
the analyzing powers $A_y$ calculated with the M3Y-P6 interaction are depicted
for the $p$-$^{40}$Ca, $^{90}$Zr and $^{208}$Pb elastic scatterings
at $\epsilon_p=16$ and $65\,\mathrm{MeV}$,
in comparison with the data.
The results of $\mathcal{V}^\mathrm{SFP}$,
$\tilde{\mathcal{V}}^{\mathrm{SFP}(\tilde{\mathrm{C}}+\mathrm{LS})}$
and $\tilde{\mathcal{V}}^\mathrm{SFP}$ are presented.
It is found that $\mathcal{V}^\mathrm{SFP}$ (\textit{i.e.}, the non-local SFP)
via M3Y-P6 successfully reproduces the data.
The agreement is comparable to the results
obtained from the empirical potentials~\cite{ref:KD03},
which are not shown here.
Analogously to $d\sigma_\mathrm{el}/d\Omega$,
we find visible deviation of the results of $\tilde{\mathcal{V}}^\mathrm{SFP}$
from those of $\mathcal{V}^\mathrm{SFP}$
at $\theta_\mathrm{c.m.}\gtrsim 50^\circ$.
The non-local SFP $\mathcal{V}^\mathrm{SFP}$
always agrees with the data better than $\tilde{\mathcal{V}}^\mathrm{SFP}$.

\begin{figure}
  \includegraphics[scale=0.7]{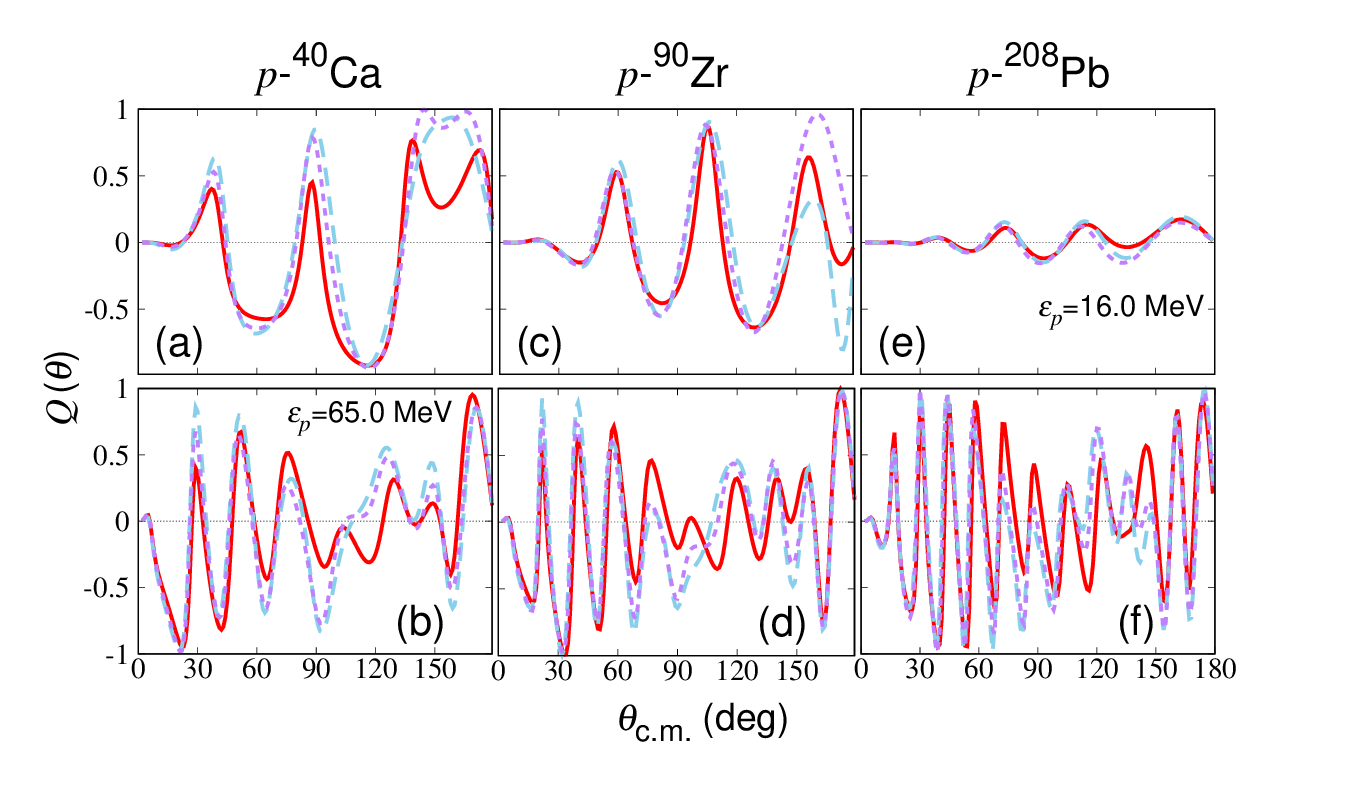}
\caption{Spin rotations of $p$-$A$ elastic scatterings:
  (a) $p$-$^{40}$Ca at $\epsilon_p=16.0\,\mathrm{MeV}$,
  (b) $p$-$^{40}$Ca at $\epsilon_p=65.0\,\mathrm{MeV}$,
  (c) $p$-$^{90}$Zr at $\epsilon_p=16.0\,\mathrm{MeV}$,
  (d) $p$-$^{90}$Zr at $\epsilon_p=65.0\,\mathrm{MeV}$,
  (e) $p$-$^{208}$Pb at $\epsilon_p=16.0\,\mathrm{MeV}$,
  (f) $p$-$^{208}$Pb at $\epsilon_p=65.0\,\mathrm{MeV}$.
  Results of $\mathcal{V}^\mathrm{SFP}$,
  $\tilde{\mathcal{V}}^{\mathrm{SFP}(\tilde{\mathrm{C}}+\mathrm{LS})}$
  and $\tilde{\mathcal{V}}^\mathrm{SFP}$
  are depicted by red solid, purple dotted and skyblue dashed lines,
  respectively.
\label{fig:p-A_Q}}
\end{figure}

Whereas no experimental data are available,
we display the results of the spin rotation $Q$ in Fig.~\ref{fig:p-A_Q}.
Although the LA almost maintains the oscillating behavior, ups and downs,
the deviation is found except at small angles.

It is found from the results
of $\tilde{\mathcal{V}}^{\mathrm{SFP}(\tilde{\mathrm{C}}+\mathrm{LS})}$
that both the LA for $v_{\alpha\beta}^{(\mathrm{C})}$
and $v_{\alpha\beta}^{(\mathrm{LS})}$ influences the spin observables,
\textit{i.e.}, the analyzing powers and spin rotations.

\subsection{Scattering off halo nucleus $^{86}$Ni\label{subsec:Ni86}}

Within the LA, the DM is approximated via Eq.~\eqref{eq:Slater}.
This approximation was confirmed to be good in the bulk
but found to be worse as the density dropped~\cite{ref:NV72}.
This raises an additional concern about the LA
for the scattering off halo nuclei,
in which the density decreases slowly.
Since the parameters in the empirical optical potential
have been determined from the data on the stable targets,
which have normal density distributions,
it is also interesting to compare the scattering observables
predicted by the SFP and the empirical potential.

\begin{figure}
  \includegraphics[scale=0.7]{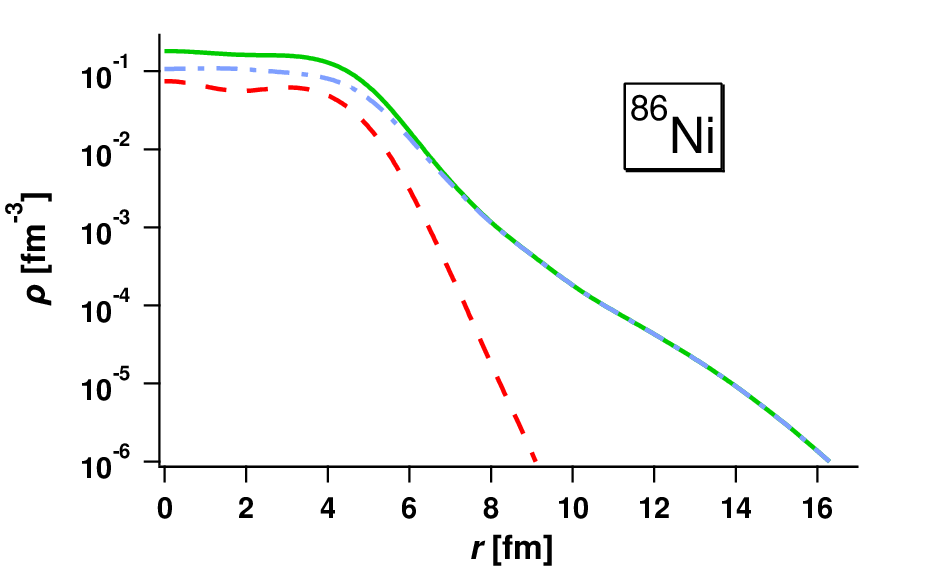}
\caption{Density profile of $^{86}$Ni
  predicted by the Hartree-Fock calculation with the M3Y-P6 interaction.
  Proton, neutron, and matter densities are shown
  by red dashed, blue dot-dashed, and green solid lines.
\label{fig:rho_Ni86}}
\end{figure}

It has been predicted that $N=58$ may behave as a magic number
at the neutron-rich nucleus $^{86}$Ni~\cite{ref:Nak10}.
The highest occupied neutron s.p. orbit is $2s_{1/2}$.
Thus, this nucleus can form a neutron halo
rationally described in the spherical HF framework.
The density profile calculated with M3Y-P6
is exhibited in Fig.~\ref{fig:rho_Ni86},
in which a neutron halo is indeed predicted.

\begin{figure}
  \includegraphics[scale=0.5]{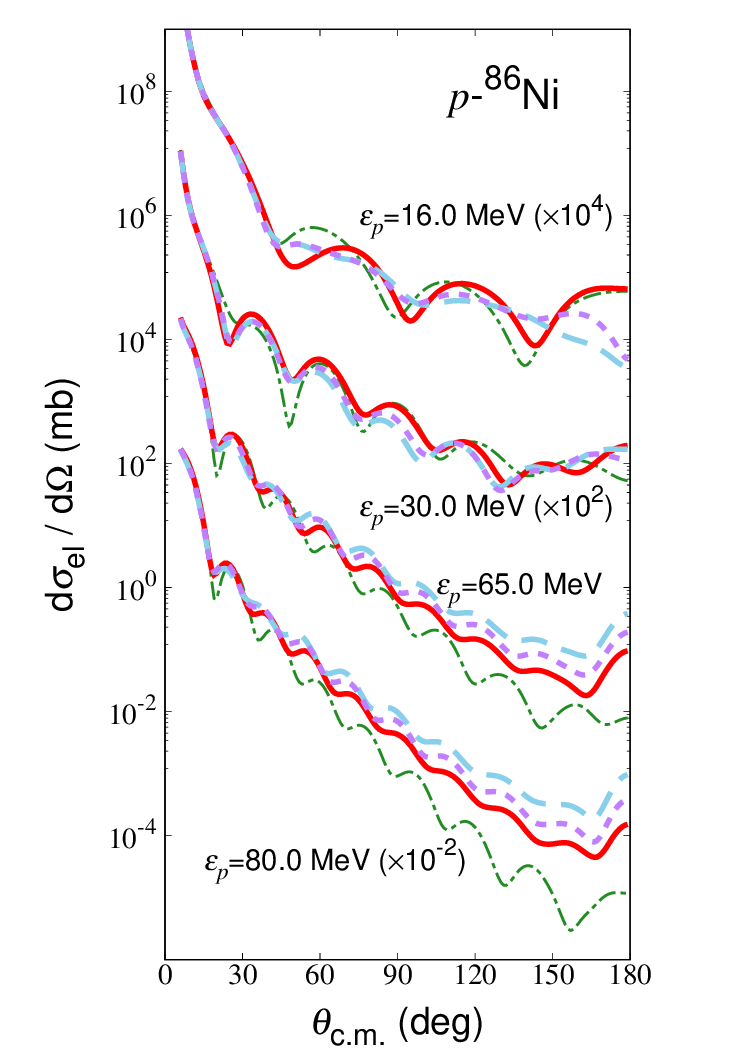}
  \caption{Predicted differential cross-sections
  of $p$-$^{86}$Ni elastic scattering
  at $\epsilon_p=16.0$, $30.0$, $65.0$ and $80.0\,\mathrm{MeV}$.
  Results of $\mathcal{V}^\mathrm{SFP}$,
  $\tilde{\mathcal{V}}^{\mathrm{SFP}(\tilde{\mathrm{C}}+\mathrm{LS})}$
  and $\tilde{\mathcal{V}}^\mathrm{SFP}$
  are depicted by red solid, purple dotted and skyblue dashed lines,
  respectively.
  For comparison, the results of the empirical potential
  in Ref.~\cite{ref:KD03}
  are also displayed by green dot-dashed lines.
  Depending on $\epsilon_p$,
  the values are scaled by the coefficient given in the parenthesis.
\label{fig:p-Ni86_dsig}}
\end{figure}

\begin{figure}
  \includegraphics[scale=0.7]{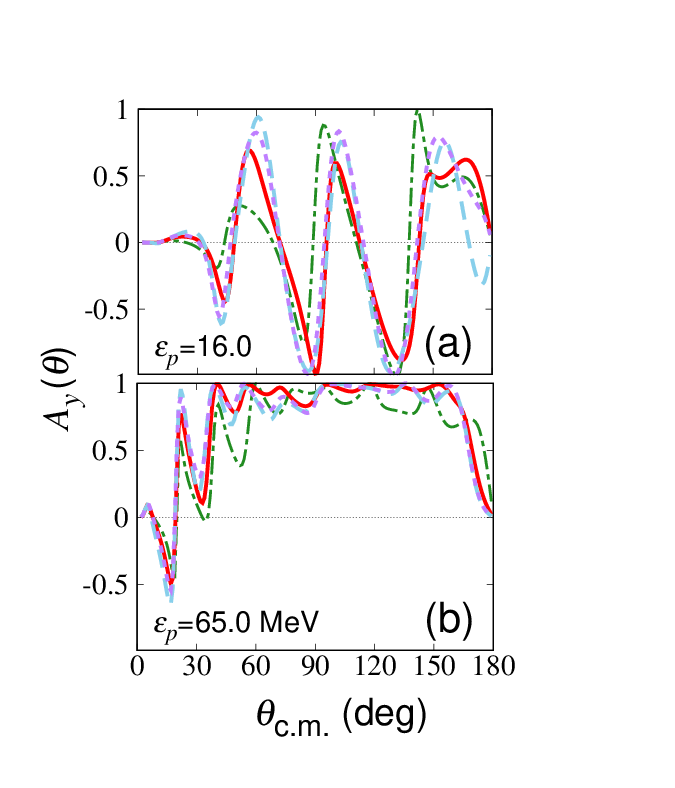}
\caption{Analyzing power of $p$-$^{86}$Ni elastic scattering
  at (a) at $\epsilon_p=16.0\,\mathrm{MeV}$
  and (b) $\epsilon_p=65.0\,\mathrm{MeV}$.
  Results of $\mathcal{V}^\mathrm{SFP}$,
  $\tilde{\mathcal{V}}^{\mathrm{SFP}(\tilde{\mathrm{C}}+\mathrm{LS})}$,
  $\tilde{\mathcal{V}}^\mathrm{SFP}$,
  and the empirical potential in Ref.~\cite{ref:KD03} are depicted
  by red solid, purple dotted, skyblue dashed, and green dot-dashed lines.
\label{fig:p-Ni86_Ay}}
\end{figure}

The observables in the proton elastic scattering off $^{86}$Ni,
$d\sigma_\mathrm{el}/d\Omega$ and $A_y$,
are computed and displayed in Figs.~\ref{fig:p-Ni86_dsig}
and \ref{fig:p-Ni86_Ay}.
On the precision of the LA,
the deviation of the $\tilde{\mathcal{V}}^\mathrm{SFP}$ results
from the $\mathcal{V}^\mathrm{SFP}$ results
is similar to the cases of the stable targets discussed above,
except at $\epsilon_p=16\,\mathrm{MeV}$.
At $\epsilon_p=16\,\mathrm{MeV}$,
the deviation starts at a smaller angle than the stable target,
and the first dip is unclear
in the $\tilde{\mathcal{V}}^\mathrm{SFP}$ result.
When the results of $\mathcal{V}^\mathrm{SFP}$ are compared
with those of the empirical potential,
we notice that the positions of the dips,
even the first dips at individual energies,
do not match well,
unlike the cases of the stable targets indicated in Ref.~\cite{ref:NI24}.
At $\epsilon_p=65$ and $80\,\mathrm{MeV}$,
the SFP yields the dips at smaller angles than the empirical potential,
which can be interpreted to reflect the large mean radius
(\textit{i.e.}, the wide density distribution) shown in Fig.~\ref{fig:rho_Ni86}.
On the contrary, the first dip in the $\mathcal{V}^\mathrm{SFP}$ result
shifts toward a larger angle at $\epsilon_p=16\,\mathrm{MeV}$.
The non-locality might influence,
as we observe that the $\tilde{\mathcal{V}}^\mathrm{SFP}$ result
of $d\sigma_\mathrm{el}/d\Omega$ has a similar structure around the first dip
to the result of the empirical potential.
The discrepancy in $A_y$ between the SFP and the empirical potential
is more apparent than the stable targets.

The virtual excitations, which arise from many-body correlations,
lead to the imaginary potential.
They may also influence the real potential $\mathcal{V}$
at the higher order of the perturbation,
called dynamical polarization (DP) effects.
Resonating with the spirit of the SCMF or the KS approaches,
significant parts of the DP effects could be incorporated
into the effective interaction,
as suggested by the success of $\mathcal{V}^\mathrm{SFP}$
for the stable targets.
However, virtual excitations to the continuum may become more significant
in halo nuclei than in stable nuclei.
Although $\mathcal{V}^\mathrm{SFP}$ takes account of the broad distribution
of the target wave function
that is ignored in the empirical potential,
we still need to care for the influence of the continuum
in the description of scattering off halo nuclei.

\section{Summary\label{sec:summary}}

We have examined the precision of the local approximation (LA)
of Refs.~\cite{ref:BR77a,ref:BR77b,ref:BR78}
on the single-folding potential (SFP).
Covering light- to heavy-mass target nuclei
and low to intermediate incident energies,
we treat the $p$-$^{16}$O, $^{40}$Ca, $^{90}$Zr and $^{208}$Pb
scatterings at $\epsilon_p=16\,-\,80\,\mathrm{MeV}$.
The target wave functions have been obtained
in the self-consistent Hartree-Fock approaches with the M3Y-P6 interaction.
The SFP (real part) is calculated consistently with the target wave functions,
applying the same M3Y-P6 interaction,
while the imaginary potential is supplemented by the empirical one.
The SIDES code~\cite{ref:SIDES} has been applied
to computations of the scattering observables,
including those under the non-local optical potential.
It was reported in Ref.~\cite{ref:NI24}
that the measured differential cross sections are reproduced,
together with the formulae for the SFP.
It is shown here that the analyzing powers are reproduced as well.

By comparing the results with and without the LA,
a sizable deviation is found at $\theta_\mathrm{c.m.}\gtrsim 50^\circ$,
although the LA works well at $\theta_\mathrm{c.m.}\lesssim 30^\circ$.
The non-local SFP is always in better agreement with the data
than the locally-approximated SFP.
The precision of the LA correlates to the momentum transfer $q$
better than $\theta_\mathrm{c.m.}$,
and the discrepancy becomes sizable at $q\gtrsim 1.5\,\mathrm{fm}^{-1}$.
Having examined the LA for the central and LS channels
of the effective interaction separately,
we find that their interplay should not be discarded.
as both influence the differential cross-sections and the spin observables.

To assess the LA in halo nuclei,
we have carried out calculations for the $^{86}$Ni target,
which is predicted to have a neutron halo
and a doubly magic nature with $N=58$
due to the occupation of the $2s_{1/2}$ orbit.
The precision of the LA in the scattering observables
is slightly worse for $^{86}$Ni than for the stable nuclei.
Since the empirical potential may not be adequate
for the spread target wave function forming the halo,
we compare the results of the SFP with those of the empirical potential as well.
Differences are found in the positions of the dips and peaks
of the differential cross sections.
Some shifts in the dip positions can be interpreted
as an effect of the large mean radius,
which is taken into account by the SFP but not in the empirical potential;
others might indicate an effect of the non-locality of the potential.

The present results will benchmark the LA on the SFP.
As assessment of the LA in the nucleus-nucleus ($A$-$A$) scatterings
is also of interest,
practical formulae and computer codes
for non-local double folding potentials are awaited.

\appendix*

\section{Summary of Brieva-Rook local approximation\label{app:LA}}

This Appendix summarizes the local approximation (LA) used in this work,
which was formulated by Brieva and Rook~\cite{ref:BR77a,ref:BR77b,ref:BR78}.
We consider the effective nucleonic interaction $v_{\alpha\beta}$
comprised of the terms given in Eq.~\eqref{eq:effint}.
The tensor channel is neglected in the LA.

Within the LA,
the SFP is decomposed into the central and $\ell s$ potentials,
\begin{equation}
  \tilde{U}_\tau^{(\mathrm{ct})}(r)
  + \tilde{U}_\tau^{(\ell s)}(r)\,\boldsymbol{\ell}\cdot\mathbf{s}\,,
\end{equation}
where $\tau=p,n$ and $\boldsymbol{\ell}=\mathbf{r}\times\mathbf{p}$.
The central channel of the interaction
$v_{\alpha\beta}^{(\mathrm{C})}+v_{\alpha\beta}^{(\mathrm{C}\rho)}$
yields central potential $\tilde{U}_\tau^{(\mathrm{ct})}(r)$
and the LS channel $v_{\alpha\beta}^{(\mathrm{LS})}$ yields
$\tilde{U}_\tau^{(\ell s)}(r)\,\boldsymbol{\ell}\cdot\mathbf{s}$.
They are related to the symbols in Sec.~\ref{subsec:LocApp}
as $\tilde{\mathcal{V}}^{(\mathrm{ct})}=\tilde{U}_\tau^{(\mathrm{ct})}(r)$
and $\tilde{\mathcal{V}}^{(\ell s)}
=\tilde{U}_\tau^{(\ell s)}(r)\,\boldsymbol{\ell}\cdot\mathbf{s}$.
The contribution of the Coulomb interaction
is included in $\tilde{U}_p^{(\mathrm{ct})}(r)$, as well.
Note that the LA induces $\epsilon_N$-dependence of
$\tilde{U}_\tau^{(\mathrm{ct})}$ and $\tilde{U}_\tau^{(\ell s)}$,
though not explicitly shown.

\subsection{Central channel\label{subapp:cent}}

The central channel $v_{\alpha\beta}^{(\mathrm{C})}$ has the form,
\begin{equation} v_{\alpha\beta}^{(\mathrm{C})}
= \sum_k \big\{t_k^{(\mathrm{SE})} P_\mathrm{SE}
+ t_k^{(\mathrm{TE})} P_\mathrm{TE} + t_k^{(\mathrm{SO})} P_\mathrm{SO}
+ t_k^{(\mathrm{TO})} P_\mathrm{TO}\big\}\,
f_k^{(\mathrm{C})} (r_{\alpha\beta})\,.
\label{eq:v-cent}\end{equation}
$P_\mathrm{Y}$ ($\mathrm{Y}=\mathrm{SE},\mathrm{TE},\mathrm{SO},\mathrm{TO}$)
stands for the projection operators on the singlet-even (SE), triplet-even (TE),
singlet-odd (SO) and triplet-odd (TO) two-nucleon states,
which are related to the spin- and isospin-exchange operators
$P_\sigma\,[=(1+4\mathbf{s}_\alpha\cdot\mathbf{s}_\beta)/2]$ and $P_\tau$ as
\begin{equation}\begin{split}
P_\mathrm{SE} = \frac{1-P_\sigma}{2}\,\frac{1+P_\tau}{2}\,,
\quad& P_\mathrm{TE} = \frac{1+P_\sigma}{2}\,\frac{1-P_\tau}{2}\,,\\
P_\mathrm{SO} = \frac{1-P_\sigma}{2}\,\frac{1-P_\tau}{2}\,.
\quad& P_\mathrm{TO} = \frac{1+P_\sigma}{2}\,\frac{1+P_\tau}{2}\,.
\end{split}\label{eq:proj_T}\end{equation}
The subscript $k$ may distinguish a parameter
(\textit{e.g.}, the range parameter)
in the function $f_k^{(\mathrm{C})}(r)$,
to which coupling constants $t_k^{(\mathrm{Y})}$ are attached.
While $f_k(r)$ is the Yukawa function in the present case,
the following discussion does not depend on  the function form.

In the coordinate representation,
the spin-independent density matrix (DM) of the target $A'$ is defined by
\begin{equation}
  \varrho_\tau(\mathbf{r}_\alpha,\mathbf{r}_\beta)
  = \sum_{\alpha,\beta\in A'} \sum_\sigma
  \varphi_\alpha^\ast(\mathbf{r}_\alpha\sigma\tau)\,
  \varphi_\beta(\mathbf{r}_\beta\sigma\tau)\,,
\end{equation}
where $\varphi_\alpha(\mathbf{r}\sigma\tau)$ is a s.p. wave function.
The local density is its diagonal part,
\begin{equation}
  \rho_\tau(\mathbf{r}) = \varrho_\tau(\mathbf{r},\mathbf{r})\,.
\end{equation}
As presented in Eq.~(24) of Ref.~\cite{ref:BR77b},
the Slater approximation yields
\begin{equation}
  \varrho_\tau(\mathbf{r}_\alpha,\mathbf{r}_\beta)
  \approx \tilde{\varrho}_\tau(\mathbf{r}_\alpha,\mathbf{r}_\beta)
  = \rho_\tau(\mathbf{R}_{\alpha\beta})\,
    \frac{3 j_1(z_{\alpha\beta})}{z_{\alpha\beta}}\,;~
    z_{\alpha\beta}
    := r_{\alpha\beta}\,k_{\mathrm{F}\tau}(\mathbf{R}_{\alpha\beta})\,,
\label{eq:Slater}\end{equation}
where $j_\lambda(z)$ is the spherical Bessel function,
with $\mathbf{R}_{\alpha\beta}=(\mathbf{r}_\alpha+\mathbf{r}_\beta)/2$,
$\mathbf{r}_{\alpha\beta}=\mathbf{r}_\alpha-\mathbf{r}_\beta$
and $k_{\mathrm{F}\tau}(\mathbf{R}_{\alpha\beta})
=\big[3\pi^2\rho_\tau(\mathbf{R}_{\alpha\beta})\big]^{1/3}$.

To simplify the expression,
we define
\begin{equation}\begin{split}
    w^{(\mathrm{C,dir/exc})}_{\tau\tau}(r_{\alpha\beta})
    &= \frac{1}{4} \sum_k \big[t_k^{(\mathrm{SE})}
      \pm 3 t_k^{(\mathrm{TO})}\big]\,f_k^{(\mathrm{C})}(r_{\alpha\beta})\,,\\
    w^{(\mathrm{C,dir/exc})}_{\tau\bar{\tau}}(r_{\alpha\beta})
    &= \frac{1}{8} \sum_k \big[t_k^{(\mathrm{SE})} + 3 t_k^{(\mathrm{TE})}
      \pm t_k^{(\mathrm{SO})} \pm 3 t_k^{(\mathrm{TO})}\big]\,
    f_k^{(\mathrm{C})}(r_{\alpha\beta})\,.
\end{split}\label{eq:w_C}\end{equation}
Here $\bar{\tau}$ stands for the counterpart of $\tau\,(=p,n)$.
As given in Eq.~(27) of Ref.~\cite{ref:BR77b},
the central part of the SFP after the LA is,
\begin{equation}\begin{split}
 \tilde{U}_\tau^{(\mathrm{ct})}(r_N)
 = \int d^3r_\beta &\Big\{\big[\rho_\tau(\mathbf{r}_\beta)\,
 w^{(\mathrm{C,dir})}_{\tau\tau}(r_{N\beta})
 + \rho_{\bar{\tau}}(\mathbf{r}_\beta)\,
 w^{(\mathrm{C,dir})}_{\tau\bar{\tau}}(r_{N\beta})\big] \\
 & + \big[\tilde{\varrho}_\tau(\mathbf{r}_N,\mathbf{r}_\beta)\,
 w^{(\mathrm{C,exc})}_{\tau\tau}(r_{N\beta})
 + \tilde{\varrho}_{\bar{\tau}}(\mathbf{r}_N,\mathbf{r}_\beta)\,
 w^{(\mathrm{C,exc})}_{\tau\bar{\tau}}(r_{N\beta})\big]\,j_0(y_{N\beta})\Big\}\,,
\end{split}\label{eq:U-cent}\end{equation}
where the scattered nucleon is expressed by the subscript $N$
instead of $\alpha$.
In the above expression,
we denote $y_{N\beta}:=r_{N\beta}\,\tilde{k}_\tau(r_N)$,
with the local momentum $\tilde{k}_\tau(r)$
defined in Eq.~(17) of Ref.~\cite{ref:BR78},
\begin{equation}
  \epsilon_N = \frac{\big[\tilde{k}_\tau(r)]^2}{2M}
  + \tilde{U}_\tau^{(\mathrm{ct})}(r)\,.
\label{eq:loc-mom}\end{equation}

\subsection{LS channel\label{subapp:LS}}

The LS channel has the form,
\begin{equation} v_{\alpha\beta}^{(\mathrm{LS})}
  = \sum_k \big\{t_k^{(\mathrm{LSE})} P_\mathrm{TE}
  + t_k^{(\mathrm{LSO})} P_\mathrm{TO}\big\}\,
  f_k^{(\mathrm{LS})} (r_{\alpha\beta})\,\mathbf{L}_{\alpha\beta}\cdot
  (\mathbf{s}_\alpha+\mathbf{s}_\beta)\,,
\label{eq:v-LS}\end{equation}
where $\mathbf{L}_{\alpha\beta}
= \mathbf{r}_{\alpha\beta}\times \mathbf{p}_{\alpha\beta}$
with $\mathbf{p}_{\alpha\beta}= (\mathbf{p}_\alpha - \mathbf{p}_\beta)/2$.
We shall use the expression,
\begin{equation}
    w^{(\mathrm{LSE})}(r_{\alpha\beta})
    = \sum_k t_k^{(\mathrm{LSE})}\,f_k^{(\mathrm{LS})} (r_{\alpha\beta})\,,\quad
    w^{(\mathrm{LSO})}(r_{N\beta})
    = \sum_k t_k^{(\mathrm{LSO})}\,f_k^{(\mathrm{LS})} (r_{\alpha\beta})\,.
\label{eq:w_LS}\end{equation}
In Ref.~\cite{ref:BR78},
the LA for the $\ell s$ part was given as
\begin{equation}\begin{split}
 \tilde{U}_\tau^{(\ell s)}(r_N)
 =& -\frac{2\pi}{3} \sum_{\tau'} B_{\tau\tau'}\,
 \frac{1}{r_N}\frac{\partial\rho_{\tau'}(\mathbf{r}_N)}{\partial r_N}\,;\\
   & B_{\tau\tau} = \int_0^\infty dr_{N\beta}\,
   w^{(\mathrm{LSO})}(r_{N\beta})\,
   r_{N\beta}^4 \big[1 + \frac{3 j_1(y_{N\beta})}{y_{N\beta}}\big] \\
   & B_{\tau\bar{\tau}} = \frac{1}{2} \int_0^\infty dr_{N\beta}
   \Big\{\big[w^{(\mathrm{LSO})}(r_{N\beta})\,
     r_{N\beta}^4 \big[1 + \frac{3 j_1(y_{N\beta})}{y_{N\beta}}\big] \\
     &\qquad\qquad + w^{(\mathrm{LSE})}(r_{N\beta})\,
     r_{N\beta}^4 \big[1 - \frac{3 j_1(y_{N\beta})}{y_{N\beta}}\big]\Big\}\,.
\end{split}\label{eq:U-ls}\end{equation}

\begin{acknowledgments}

This work is supported by the JSPS KAKENHI, Grant No. JP24K07012.
A part of the numerical calculations has been performed on HITAC SR24000
at Institute of Management and Information Technologies in Chiba University.
\end{acknowledgments}


\begin{thebibliography}{99}
\bibitem{ref:Gle83} N.K.~Glendenning, \textit{Direct Nuclear Reactions}
  (Academic Press, New York, 1983).
\bibitem{ref:Sat83} G.R.~Satchler, \textit{Direct Nuclear Reactions}
  (Oxford Univ. Press, Oxford, 1983).
\bibitem{ref:SL79} G.R.~Satchler and W.G.~Love,
  Phys. Rep. \textbf{55}, 183 (1979).
\bibitem{ref:BR77a} F.A.~Brieva and J.R.~Rook,
  Nucl. Phys. A \textbf{291}, 299 (1977).
\bibitem{ref:BR77b} F.A.~Brieva and J.R.~Rook,
  Nucl. Phys. A \textbf{291}, 317 (1977).
\bibitem{ref:BR78} F.A.~Brieva and J.R.~Rook,
  Nucl. Phys. A \textbf{297}, 206 (1978).
\bibitem{ref:CH89} R.L.~Varner, W.J.~Thompson, T.L.~McAbee, E.J.~Ludwig
  and T.B.~Clegg, Phys. Rep. \textbf{201}, 57 (1991).
\bibitem{ref:KD03} A.J.~Koning and J.P.~Delaroche,
  Nucl. Phys. A \textbf{713}, 231 (2003).
\bibitem{ref:LKP20} D.T.~Loan, D.T.~Khoa and N.H.~Phuc,
  J. Phys. G \textbf{47}, 035106 (2020).
\bibitem{ref:ABL90} H.F.~Arellano, F.A.~Brieva and W.G.~Love,
  Phys. Rev. C \textbf{42}, 652 (1990).
\bibitem{ref:AB24} H.F.~Arellano and G.~Blanchon,
  Phys. Rev. C \textbf{109}, 064609 (2024).
\bibitem{ref:M3Y} G. Bertsch, J. Borysowicz, H. McManus and W.G. Love,
  Nucl. Phys. A \textbf{284}, 399 (1977).
\bibitem{ref:M3Y-P} N. Anantaraman, H. Toki and G.F. Bertsch,
  Nucl. Phys. A \textbf{398}, 269 (1983).
\bibitem{ref:Nak03} H.~Nakada, Phys. Rev. C \textbf{68}, 014316 (2003).
\bibitem{ref:Nak13} H.~Nakada, Phys. Rev. C \textbf{87}, 014336 (2013).
\bibitem{ref:NS14} H.~Nakada and K.~Sugiura,
  Prog. Theor. Exp. Phys. \textbf{2014}, 033D02;
  \textit{ibid} \textbf{2016}, 099201.
\bibitem{ref:Nak20} H.~Nakada,
  Int. J. Mod. Phys. E \textbf{29}, 1930008 (2020).
\bibitem{ref:NI24} H.~Nakada and K.~Ishida,
  Phys. Rev. C \textbf{109}, 044614 (2024).
\bibitem{ref:SIDES} G.~Blanchon, M.~Dupuis, H.F.~Arellano, R.N.~Bernard
  and B. Morillon,
  Comp. Phys. Comm. \textbf{254}, 107340 (2020).
\bibitem{ref:RKG84} L.~Rikus, K.~Nakano and H.V.~von~Geramb,
  Nucl. Phys. A \textbf{414}, 413 (1984);
  L.~Rikus and H.V.~von~Geramb, Nucl. Phys. A \textbf{426}, 496 (1984).
\bibitem{ref:YNM86} N.~Yamaguchi, S.~Nagata and J.~Michiyama,
  Prog. Theor. Phys. \textbf{76}, 1289 (1986).
\bibitem{ref:ADG00} K.~Amos, P.J.~Dortmans, H.V.~Von~Geramb, S.~Karataglidis
  and J.~Raynal, in \textit{Adv. Nucl. Phys.} vol.~25,
  edited by J.W.~Negele and E.~Vogt (Plenum, New York, 2000), p.~275.
\bibitem{ref:KS65} W.~Kohn and L.J.~Sham, Phys. Rev. \textbf{140}, A1133 (1965).
\bibitem{ref:Nak23} H.~Nakada, Phys. Scr. \textbf{98}, 105007 (2023).
\bibitem{ref:NS06} H.~Nakada and T.~ Shinkai, arXiv:nucl-th/0608012.
\bibitem{ref:CB78} X.~Campi and A.~Bouyssy,
  Phys. Lett. B \textbf{73}, 263 (1978).
\bibitem{ref:NV72} J.W.~Negele and D.~Vautherin,
  Phys. Rev. C \textbf{5}, 1472 (1972).
\bibitem{ref:DCK10} J.~Dobaczewski, B.G.~Carlsson and M.~Kortelainen,
  J. Phys. G \textbf{37}, 075106 (2010).
\bibitem{ref:VB72} D.~Vautherin and D.M.~Brink,
 Phys. Rev. C \textbf{5}, 626 (1972).
\bibitem{ref:Gogny} J.~Decharg\'{e} and D.~Gogny,
 Phys. Rev. C \textbf{21}, 1568 (1980).
\bibitem{ref:SLKR95} M.M.~Sharma, G.~Lalazissis, J.~K\"{o}nig and P.~Ring,
 Phys. Rev. Lett. \textbf{74}, 3744 (1995).
\bibitem{ref:NI15} H.~Nakada and T.~Inakura,
 Phys. Rev. C \textbf{91}, 021302(R) (2015).
\bibitem{ref:Nak15} H.~Nakada, Phys. Rev. C \textbf{92}, 044307 (2015).
\bibitem{ref:Nak19} H.~Nakada, Phys. Rev. C \textbf{100}, 044310 (2019).
\bibitem{ref:Nak10} H.~Nakada, Phys. Rev. C \textbf{81}, 051302(R) (2010).
\bibitem{ref:JLM76} J.P.~Jeukenne, A.~Lejeune and C.~Mahaux,
  Phys. Rep. \textbf{25}, 83 (1976);
  E.~Bauge, J.P.~Delaroche and M.~Girod,
  Phys. Rev. C \textbf{63}, 024607 (2001).
\bibitem{ref:FSY08} T.~Furumoto, Y.~Sakuragi and Y.~Yamamoto,
  Phys. Rev. C \textbf{78}, 044610 (2008).
\bibitem{ref:HKMW13} J.W.~Holt, N.~Kaiser, G.A.~Miller and W.~Weise,
  Phys. Rev. C \textbf{88}, 024614 (2013).
\bibitem{ref:VFG16} M.~Vorabbi, P.~Finelli and C.~Giusti,
  Phys. Rev. C \textbf{93}, 034619 (2016).
\bibitem{ref:WLH21} T.R.~Whitehead, Y.~Lim and J.W.~Holt,
  Phys. Rev. Lett. \textbf{127}, 182502 (2021).
\bibitem{ref:MO12} K.~Mizuyama and K.~Ogata,
  Phys. Rev. C \textbf{86}, 041603(R) (2012).
\bibitem{ref:Bla15} G.~Blanchon, M.~Dupuis, H.F.~Arellano and N.~Vinh~Mau,
  Phys. Rev. C \textbf{91}, 041612 (2015).
\bibitem{ref:EXFOR} https://www-nds.iaea.org/exfor/.
\bibitem{ref:exp_DCS_RCNP} H.~Sakaguchi, M.~Nakamura, K.~Hatanaka, A.~Goto,
  T.~Noro, F.~Ohtani, H.~Sakamoto, H.~Ogawa and S.~Kobayashi,
  Phys. Rev. C \textbf{26}, 944 (1982).
\bibitem{ref:exp_DCS_p-O16_e17} G.M.~Crawley and G.T.~Garvey,
  Phys. Rev. \textbf{160}, 981 (1967).
\bibitem{ref:exp_DCS_p-O16_e30} P.D.~Greaves, V.~Hnizdo, J.~Lowe and O.~Karban,
  Nucl. Phys. A \textbf{179}, 1 (1972).
\bibitem{ref:exp_DCS_p-O16_e50} J.A.~Fannon, E.J.~Burge, D.A.~Smith
  and N.K.~Ganguly,
  Nucl. Phys. A \textbf{97}, 263 (1967).
\bibitem{ref:exp_DCS_p-A_e40} L.N.~Blumberg, E.E.~Gross, A.~Van~Der~Woude,
  A.~Zucker and R.H.~Bassel,
  Phys. Rev. \textbf{147}, 812 (1966).
\bibitem{ref:exp_DCS_p-A_e80} A.~Nadasen, P.~Schwandt, P.P.~Singh, W.W.~Jacobs,
  A.D.~Bacher, P.T.~Debevec, M.D.~Kaitchuck and J.T.~Meek,
  Phys. Rev. C \textbf{23}, 1023 (1981).
\bibitem{ref:exp_DCS_p-Pb208} W.T.H.~van~Oers, H.~Haw, N.E.~Davison,
  A.~Ingemarsson, B.~Fagerstr\"{o}m and G.~Tibell,
  Phys. Rev. C \textbf{10}, 307 (1974).

\end{thebibliography}

\end{document}